# Reclaiming Epistemic Agency: A Critical Framework for Human-Generative AI Co-Agency in Education

Biranchi Poudyal
Faculty of Arts and Society, Charles Darwin University, Australia
ORCID: https://orcid.org/0000-0002-7210-5480

## Abstract

Generative artificial intelligence (GenAI) has been primarily framed as an impartial educational tool. However, this framing overlooks an even larger shift: the reassignment of epistemological authority from teachers to students to machines. This paper presents a conceptual evaluation of the extent to which GenAI redistributes students' and teachers' ability to act in classrooms to produce knowledge, validate each other's claims, and create evidence of student learning while collaborating with and competing against humans. This evaluation will draw on various theoretical paradigms, including Distributed Agency, Self-Determination Theory, Society 5.0, and Technology Integration Paradigms (including, but not limited to, TPACK and SAMR). While all of the theoretical paradigms evaluated are relevant to the role of agency within education mediated by AI, none of them address the ongoing disparity regarding equitable distribution of power, ownership of the data used to mediate interaction, and accountability in relation to human-mediated interactions. As such, this paper introduces the Ecological Co-Agency Framework, which defines agency in terms of relational, regulatory, and pedagogical processes, conditioned by a defined commitment to human accountability for epistemological claims.

**Keywords:** generative AI; co-agency; epistemic authority; distributed agency; self-regulated learning; higher education

## 1. Introduction

The GenAI platform has fundamentally changed how both students and teachers develop and evaluate knowledge in educational settings through GenAI-based systems. The primary reason is that these systems enable customized approaches to learning and new collaborative methods for creating and developing knowledge. Furthermore, the GenAI platform provides many ways for students to express their creativity at scales previously impossible (Kong & Yang, 2024; Lee et al., 2025). However, while AI-based learning platforms offer many exciting new avenues for knowledge production and accessibility, they also introduce significant challenges. For example, overreliance on AI-generated content will become problematic. Additionally, there may be a greater disparity than ever before in "power" between human-created and machine-created content. Finally, about ethics, this new platform introduces a variety of new questions regarding such things as authorship, accountability, etc. (Bueno et al., 2025; Satyanarayan & Jones, 2024)

The role of "agency" is central to the transformation described in this research. Historically, agency has been viewed as an aspect of human-centeredness. For example, for many years, educators served as agents, assisting students in acquiring knowledge (Leijen et al., 2022). However, the use of GenAI technology has introduced ambiguity into the definition of agency. Although GenAI systems can produce content, evaluate the quality of the content they generate, and adapt their generated content based on the results of their evaluations without direct human involvement, the question exists as to whether machines can be considered as acting with agency at the same level as humans (Riemer & Peter, 2024). On the other hand, GenAI systems participate in the evaluation process and therefore also act within the context of the interpretation of the agency they represent. The result of including GenAI in this process is not simply a new educational resource for educators; rather, it is a redistribution of agency from human entities to non-human entities. This redistribution of agency will be referred to throughout this document as co-agency (Godwin-Jones, 2024).

While most of the empirical research about GenAI in education has examined how it affects student interest and participation, teacher workload and the validity of assessments (Johnston et al., 2024) - with GenAI's epistemological role seen as less important than issues relating to the usability and adoption of GenAI - very little research addresses the primary implications of co-agency: that when a system can create legitimate, coherent, and often persuasive content, human judgment is subjected to a new type of pressure that no amount of improvement in digital literacy or skill at writing prompts will alleviate. Thus, this study aims to fill that gap.

Using the identification of how Generative AI changes student and teacher agency, the researcher developed a theoretical model for integrating Generative AI into education while maintaining the epistemological basis on which education exists and is valued. The research will answer two questions. First, in what way(s) do the generative AI tools affect the agency of both students and teachers in their learning/teaching environments? Second, what are some conceptual and practical mechanisms that would enable an equitable and responsible incorporation of Generative AI into instruction and learning? Recognizing that agency is currently being distributed across humans and AI systems does not imply equality. There is a significant difference in how humans and AI systems generate meaningful output. When one conflates the two, a high price is paid in terms of pedagogy and knowledge.

## 2. Conceptual and Methodological Approach

The focus here was on conducting a significant conceptual research effort rather than an empirical one. Conceptual efforts are considered highly valued and distinct scholarly contributions within the social sciences and education. Unlike empirical research that generates new data from observations, surveys, etc., conceptual research synthesizes, organizes, and builds upon established theory to produce a new conceptual framework for understanding a particular issue (Jaakkola, 2020). In line with Jaakkola's (2020) classification of conceptual contributions into three types, this research will take what he calls a “model” type: It will utilize previously developed theoretical frameworks (distributed and relational agency; self-determination theory; self-regulated learning; and models of technology integration which have been widely adopted), to build a completely original conceptual model, as opposed to testing or building upon an already established prior theoretical framework.

The literature on which this study is based has been identified through a series of systematic searches conducted over time and described as ‘purposeful’ (and therefore selective) and ‘iterative’, using four databases: Scopus, Web of Science, ERIC, and Google Scholar. These searches used the citation chain methodology (a technique that identifies subsequent citations of an earlier paper) to identify other relevant studies (Bueno et al., 2025; Godwin-Jones, 2024; Kong & Yang, 2024). Given the theoretically fragmented nature of the literature under review and its rapid expansion, citation chaining is a suitable approach to synthesizing evidence in this field (Grant & Booth, 2009). In contrast to the inclusion/exclusion criteria commonly employed in reviews (e.g., NICE Technology Appraisal Guidance: Methods Guide), no formal inclusion/exclusion criteria or quality appraisal tools were applied during the selection process. Instead, references were identified for their potential to contribute to the understanding of issues related to agency, epistemic authority, and the integration of AI into learning environments. A preference has been given to including literature published post-2018, although reference will also be made to some influential pre-2018 foundational theories (Bandura, 2006; Ryan & Deci, 2000; Zimmerman, 2002) and influential trade literature (Mollick, 2024; Harari, 2024). Such selectivity limits the generalizability of the findings presented here. While the synthesis does not claim to replicate the level of replicability afforded by a PRISMA-style review, nor can it provide primary empirical data, it seeks to synthesize the rapidly evolving body of knowledge on the use of GenAI in learning environments. It also seeks to establish the basis upon which the empirical validation suggested in Section 6 may occur. As such, it is written from a perspective that is both positively disposed towards GenAI’s capacity to support instruction and sceptical of its universal benefits.

## 3. Literature Review

### 3.1 Learner and Teacher Agency After GenAI

As a twenty-first-century model based on “the four C’s” - communication, collaboration, critical thinking, and creativity, student competences have been defined for some time by a model that GenAI has prompted a re-evaluation of each through the lens of learner agency (Türker & Öztürk, 2024). The act of communication will include determining whether to delegate aspects of it to an AI program or retain them, because while a machine can produce surface-level text that may be empathetic and sensitive to its audience, the empathetic and audience-sensitive nature remains fundamentally human (Holmes et al., 2019).
Collaboration shifts from organizing tasks to negotiating with one another when using AI, resulting in one team member being reliant on the other members, rather than all members having equal responsibility for the task. The ability to think critically will involve analysing AI output in terms of both accuracy and fluency. It will no longer be enough to generate ideas.
Creativity is frequently identified as the last bastion of human thought processes. However, this characterization underestimates the extent to which creative practices are mediated by both technological tools and cultural contexts, rather than solely in individual minds (Bueno et al., 2025; Zhou & Lee, 2024). Research on AI-assisted creativity suggests that the greatest benefits arise from interfaces that allow learners to explore ideas rather than generate them themselves (O'Toole & Horvát, 2024).

In contrast to the view that the emergence of AI will make humans obsolete, many recent controlled studies have shown that the use of AI is likely to increase individual assessments of their own creativity; however, it also increases similarity in students' creative work, therefore increasing students' individual novel thinking but reducing the number of ways they think creatively in comparison to other students. Therefore, according to Doshi & Hauser (2024), an individual's enhanced creativity may ultimately reduce the amount of unique thinking present within the class. The cost of increased individualized creativity driven by AI development has been amplified by concerns about how AI research is funded, in part through the unpaid labor of marginalized communities (Colón-Vargas, 2025).

Accordingly, Mamonov (2024) posits that creativity should be viewed as the ability to guide creative processes rather than as access to the tools used in them. The three components identified by Ryan and Deci (2000) are key to understanding Self-Determination Theory. The three components include Autonomy, Competence, and Relatedness. These components drive motivation. While GenAI changes the circumstances under which Autonomy, Competence, and Relatedness occur, it does not eliminate their necessity. A student's sense of competence develops when they learn to evaluate their own AI-generated outputs. Additionally, teaching students to assess the quality of AI outputs represents an additional competency students require to develop (Lai & Bower, 2020). Similarly, developing a sense of Relatedness in a GenAI-mediated learning environment involves collaborating with peers and educators to define the ethics of AI use and determine acceptable creative boundaries for AI-generated content (Redeker, 2017). Bandura's (2006) Social-Cognitive Theory explains how individuals exert control over their behaviour through the interplay among behaviour, environmental influences, and person-based characteristics. This theory describes how an individual exhibits agency based upon an interplay between their internal characteristics, their environment, including platforms and algorithms that facilitate their interactions with those environments, and their behaviors.

**3.2 Contested Intelligence**

The debate over how "intelligence" applies to General Artificial Intelligence (Gen AI) in education is unfolding along two dimensions. Mollick (2024) describes the co-productive process of generating outputs through joint cognitive processes by humans and machines when interacting as 'co-intelligence' - rather than outputs generated solely by the processing capacity of an individual's or a machine's cognitive abilities. Harari (2024), however, creates a stark contrast between intelligence, defined as information-processing capabilities that computer systems can scale, and the subjective awareness that includes self-awareness, emotional responses, and sensory input required for consciousness, all qualities unique to humans.

Therefore, if Harari's distinction proves valid, it follows that, regardless of technological scalability, no amount of technical scale will allow the delegation of moral judgments to a system devoid of subjective interest in the outcomes of those decisions. Adaptive technology that adjusts students' curriculum pathway(s) in real time, typically referred to as 'intelligent tutoring', can also directly limit student autonomy to the extent that recommendations are treated at face value. In this manner, adaptive technology that filters information to support student decision-making may act as a type of remote control, and such are not two different technologies, but rather two differing operational configurations of the same technology (Catena et al., 2026).

### 3.3 Society 5.0 and the Politics of Human-AI Symbiosis

A second area of study places GenAI in the context of Society 5.0 - a Japanese government's view of a "human-centered" society in which the lines between the digital and physical worlds are blurred and used to solve the many complex social challenges facing the world today (Fukuyama, 2018), primarily created for industry use, but being used in education through analogy. On the other hand, Harari (2024) posits that GenAI presents a break in the way societies have distributed epistemological power: previous technological innovations increased human capabilities while maintaining human judgment; however, GenAI creates and assesses knowledge claims autonomously and will eventually displace the assessment of learner knowledge from educators unless educators intentionally monitor the evaluation process of algorithms. While there are certainly differing views on this issue, they can be considered complementary. Harari provides a cautionary note against the unrealistic expectations of Society 5.0, while Society 5.0 provides a model for designing a future state.

Ultimately, whether GenAI-based tutoring will provide personalized learning at scale (Kong & Yang, 2024; Rutherford et al., 2025) or lead students down the path of passive consumption will depend on the defaults developers embed in the system during the rollout phase. Teacher judgment regarding their role in GenAI tutoring will depend on the defaults established by developers (Zhai, 2025). Therefore, realizing Society 5.0's equilibrium will require viewing algorithmic bias, transparency, and intellectual property ownership as prerequisites for incorporating GenAI rather than afterthoughts (Fu & Weng, 2024). It is important to understand that this is not a hypothetical concern: systems trained on common linguistic and cultural inputs automatically default to Western curricula and climate assumptions regardless of input, limiting access to indigenous languages and charging rates that reduce accessibility in low-income areas in developing countries (Nyaaba et al., 2026).

### 3.4 The Proliferation of Integration Frameworks: TPACK, SAMR, and Their Limits

Existing approaches to integrating GenAI into education primarily rely on established models of technology integration. Kong & Yang (2024) developed an application model centered on the “Self-regulation” paradigm, based on TPACK (Mishra & Koehler, 2006); however, the emphasis on developing teacher professional development may be difficult to achieve with the limited resources available to today’s educators. In response, Petko et al. (2025) proposed a TPACK-based extension (with an additional ‘Contextual knowledge’ (XK) layer) as an approach to differentiate teacher knowledge regarding the use of GenAI tools and their usage within their respective educational institutions. This same differentiation was noted again in section 4.3. A second perspective on this topic is SAMR (Substitution-Augmentation-Modification-Redefinition), which offers a different lens for examining the extent of GenAI implementation in education. While SAMR has been employed as a theoretical model to provide a structure to assess the level of adoption of new technologies, Drugova et al. (2021) reported that the majority of GenAI educational applications have utilized the lowest levels of SAMR, i.e., substitution; thereby, only replicating current teaching practices without attempting to transform them.

Student interaction with GenAI has been categorized into four types by Yang et al. (2024): resistive, receptive, resourceful, and reflective. Only the latter two categories will increase student agency. Zhai (2025), like Yang et al. (2024), identified three stages of teacher involvement with GenAI: Observer, Adopter/Collaborator/Innovator; again, noting that each stage will depend on institutional support rather than individual educator inclination. Finally, Roe & Perkins (2024) conducted a scoping review of 10 studies on the potential benefits and drawbacks of using GenAI to augment traditional educational methods. The authors concluded that GenAI has the potential to personalize education and increase accessibility for students while simultaneously reducing learner involvement in problem-solving processes; therefore, they argued that learner-centered safeguards should be incorporated into all GenAI designs rather than treated as options or afterthoughts. It is worth noting that, given the relatively small number of studies included in the review, Roe & Perkins’s (2024) arguments present a compelling case for the inclusion of learner-centered safeguards in GenAI designs but do not necessarily constitute a conclusive empirical finding.

None of the reviewed literature treated GenAI integration as a primary issue of agency, nor did it discuss the power dynamics, data management issues, and ethical responsibilities inherent in GenAI that differ significantly from those associated with technologies for which TPACK and SAMR were originally designed. As such, there is a need for a framework that integrates concepts of agency, self-regulation, and pedagogical roles, along with explicitly defined ethical boundaries, as discussed in Section 4.

## 4. Towards an Ecological Framework of Co-Agency

The evidence that has been gathered from the prior research appears to support three common themes: GenAI-mediated learning is mediated by an agent other than the learner themselves (Bueno et al., 2025; Godwin-Jones, 2024); GenAI-mediated learning occurs through a regulatory process in which the learner plans, acts, and reflects (Zimmerman, 2002); and, there are different forms of GenAI-mediated learning dependent upon how teachers view technology and therefore how they position themselves professionally (Zhai, 2025). Where current theoretical models have failed is in providing a means to integrate all three dimensions and to explicitly articulate the ethically justifiable conditions under which those dimensions can be used together. This was addressed by developing the Ecological Co-Agency Model, as shown in Figure 1. It treated co-agency as the relationship between three interdependent and co-constituted dimensions - relational, regulatory, and pedagogical - within a non-negotiable boundary of human epistemological accountability.

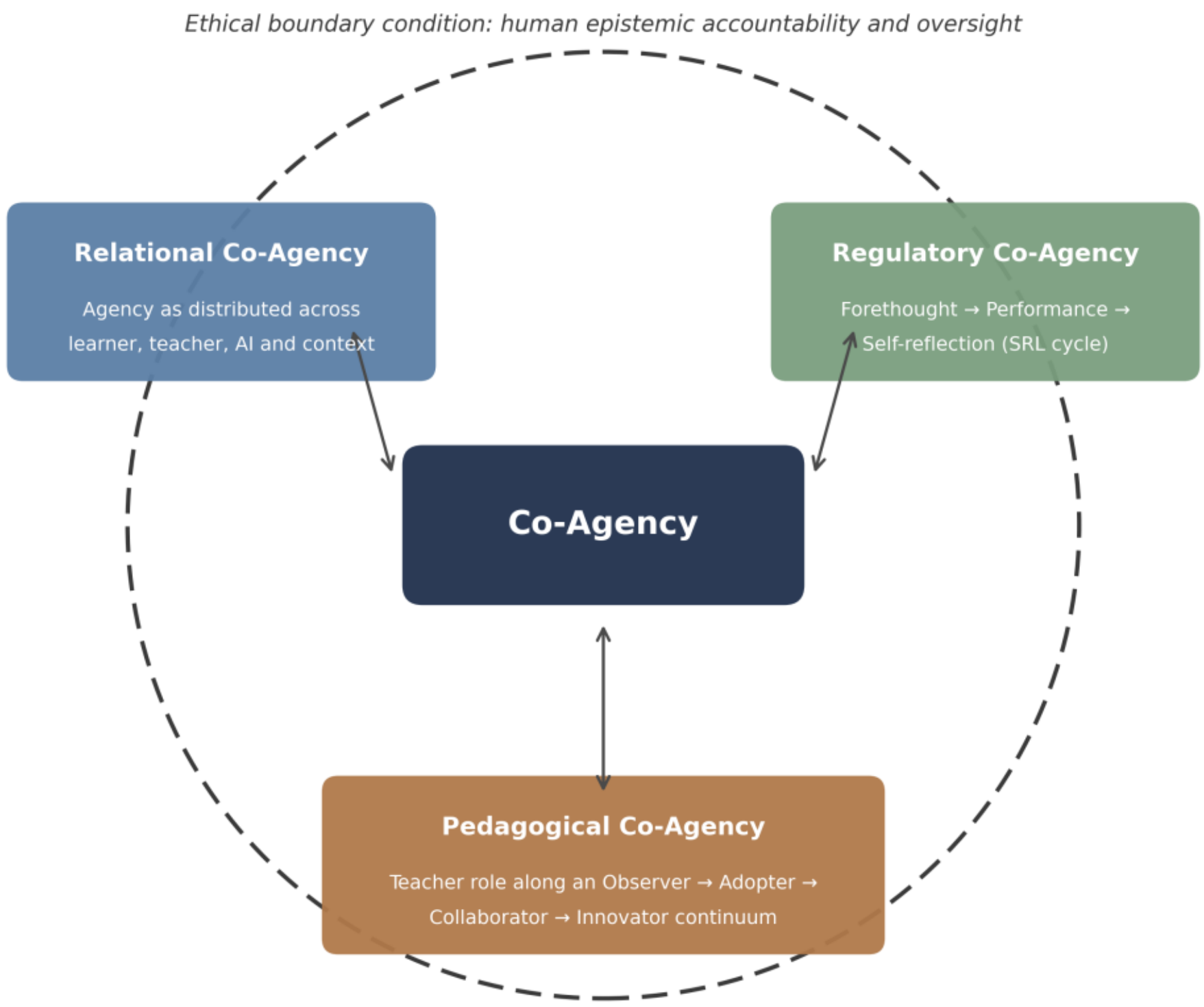

*Figure 1. The Ecological Co-Agency Framework. Relational, regulatory, and pedagogical co-agency interact bidirectionally, all bounded by the ethical condition of human epistemic accountability and oversight.*

### 4.1 Relational Co-Agency

The first perspective defines agency as resulting from interaction rather than being independent of both the person and the machine (Bueno et al., 2025). Relational Co-Agency is also based on social material views of how people learn. Social materials view competency, creativity, and judgement as resulting from the relationships formed among the student, the tool (such as technology), and the learning context, rather than as products of the student alone (Godwin-Jones, 2024). Two requirements must be met to develop relational co-agency in a learning environment. First, there needs to be a transparent description of who has made the decision for which tasks have been assigned to an AI system and which to the students. Second, relational co-agency can only occur if ethical co-agency exists. Ethical Co-Agency refers to a joint responsibility for an educational outcome between a human and an AI. In other words, while humans will still have primary responsibility and accountability for all educational outcomes generated by AI, they will act in a monitoring capacity and be accountable for how they use AI-generated outputs (Satyanarayan & Jones, 2024).

### *4.2 Regulatory Co-Agency*

To begin with, Zimmerman's (2002) theory of self-regulated learning consists of three phases of self-regulation: forethought, performance, and reflection. This framework was mapped onto an artificial-intelligence-mediated environment to create the second dimension. In the forethought phase, for example, GenAI can assist students in formulating objectives and finding sources, but students should always be in control of their objectives. Furthermore, although students may utilize GenAI as a source of choices for resources and alternatives, they must never be forced to act on behalf of GenAI.

In the performance phase, GenAI will offer feedback and prompts to students as they respond. "Augmented Intelligence" or "Scaffolding," as referred to by Dede (2010), is a type of GenAI designed to liberate cognitive space, enabling students to engage in more complex tasks while still completing them. Finally, in the reflection phase, GenAI will produce summary reports of how students have progressed throughout the course materials; however, when it is time for students to evaluate their own development and decide upon how they would like to proceed, both students and teachers are expected to do so based on data produced by GenAI. Classification systems informed by Self-Determination Theory currently categorize student activity with ChatGPT across the three phases. Such classification systems demonstrate how GenAI can be used to design specific task types that enhance regulatory agency in students (Chiu, 2024).

Offloading does not represent the antithesis of being engaged. Research employing mixed methods has demonstrated a paradoxical association between offloading and engagement across 3 cultures. Strategic cognitive off-loading, for example, involves providing a well-suited subtask to allocate cognitive capability to problems that require human judgment. Strategic cognitive offloading can lead to higher levels of transformative learning; however, this occurs only when the offloading decision is made intentionally rather than routinely (Wang & Zhang, 2026). It is in the regulation aspect of engagement that a system can either facilitate 'resourceful' and 'reflective' forms of engagement (Yang et al., 2024) or prevent such engagement.

### **4.3 Pedagogical Co-Agency**

The third dimension represents the teacher at the heart of the integration process since their evolving function is contained within a space formed by both Zhai's (2025) observer-adopter-collaborator-innovator continuum and the TPACK/XK as defined by Petko et al. (2025). A teacher can move either up or down this continuum based on how much support they receive in terms of time, training, technology, and administrative backing to expand their abilities to go beyond simply using AI to replace one of the many tasks they perform today (e.g. the "Adopter" position) to create new methods of performing those tasks (i.e. the "Innovator" position) (Drugova et al., 2021; Mishra et al., 2023).

As a result, this dimension serves as structural accountability for schools, rather than solely something that each teacher needs to develop themself: If no structural support is provided to the teacher, then the continuum of observer - adopter - collaborator - innovator becomes far less of a developmental tool for teachers and more of a means for identifying those teachers who already possess the necessary resources to experiment.

Another description provides an alternative illustration of the identical development process utilizing the terminology of 'team' as opposed to 'stages of adoption', while also describing why, if a teacher is incapable of transitioning from transactional forms of cooperative work with AI to synergistic forms of cooperative work with AI, they will suffer from cognitive atrophy and de-professionalize their own profession (Cukurova et al., 2025). This is not hypothetical. When asked in interviews how they use GenAI daily, teachers report having been assigned significant amounts of what would appear to be invisible labor related to reviewing, editing, and often rewriting AI-generated materials so that they meet classroom standards. This is an example of a teacher exercising professional judgment at the precise moment when AI is incapable of doing the same (Selwyn et al., 2025).

### 4.4 The Ethical Boundary Condition

The three dimensions illustrated above alone will not give us the information needed to judge whether GenAI will have positive or negative impacts. As stated before, even if GenAI can produce high-quality outputs, it could also create bias against certain groups and reduce human responsibility for their own decision-making. So, the framework views "human epistemic accountability" as a boundary condition to each of the three other dimensions - a limitation for all three, but one of the four dimensions.

There are three ways in which this is true: first, contestability. For learners and teachers to evaluate and trust AI-generated results, they must be able to question and/or cross-check them effectively. To accomplish this, we want some degree of explainability from the GenAI systems currently being implemented (Khosravi et al., 2022) -secondly, provenance. When institutions implement GenAI tools, they should be able to describe, in simple terms, to students the source(s) of the training data used by the tool and how that data likely introduced bias into the tool. Similarly, when institutions present the AI-generated output as epistemically neutral (Tao et al., 2024; Pragya, 2025), we need a similar standard to be applied. Thirdly, non-delegation of moral and intellectual credit. We cannot allow human decisions related to student welfare, academic standing, or opportunity (e.g., final grades, disciplinary judgments, high-stakes comments/feedback) to be determined solely by GenAI's contributions to these outcomes (Holzinger et al., 2025; Bearman et al., 2024). Authorship questions are addressed similarly. Treating an AI system as a co-author/co-creator of an academic or artistic work reduces the accountability that authorship is meant to imply (Bao & Zeng, 2026).

**4.5 Illustrating the Framework**

The three possible forms of co-agency can be illustrated through an example. A group of undergraduate students is tasked with creating a literature review and is given access to a generative artificial intelligence (GenAI) writing assistant. Relational co-agency would involve the assignment prompt specifying which tasks could be delegated to the AI (e.g., summarizing individual sources) and which could not (e.g., creating an argument linking all the individual sources to the research question). As such, the division of labor among the students would be specified by the assignment prompt, rather than each student independently determining how best to divide the workload.

Regulatory co-agency would require students to specify a thesis statement and a brief outline of their intended paper before using AI. Students would then use the AI tool to receive feedback on the structural aspects of their arguments, rather than on sentence structure, while creating the draft. Once students complete their drafts, they must also submit a very short self-reflective commentary detailing where they agreed with the suggested changes, modified those suggestions, or rejected them, and explaining why they took that position.
Pedagogical co-agency will represent a fundamental transformation in how instructors evaluate students' written products. Rather than treating the completed essay as a product to be evaluated solely on content and grammar, instructors will need to consider the "decision trail" evident throughout the writing process. Institutions will need to dedicate additional resources to transform their evaluation systems from assessing student work based on completed essays toward evaluating the decision-making processes students employed in producing their work.

Ultimately, at no time does accountability fall on the AI. The instructor retains sole responsibility for determining whether a student's work meets the minimum expectations. At all times, however, the AI tool assists in producing student work. However, unlike other tools, these contributions are transparent and subject to examination. Additionally, unlike other tools, there is no expectation that the AI tool's contributions will become indistinguishable in an unexamined final product.

## 5. Discussion: Implications for Policy and Practice

The proposed model requires a paradigmatic shift in both parties' understanding of their roles and responsibilities. Students become creators of knowledge, working collaboratively with AI to discover new ideas rather than solely relying on AI for predetermined answers (Bueno et al., 2025). Teachers transition from being the sole providers of knowledge to creating and implementing methods that enable students to learn to use GenAI responsibly (Kong & Yang, 2024). This is a major transition. Further, the model anticipates that this will not happen naturally when GenAI is first implemented. There will need to be intentional efforts by educators and school systems to design assignments that reflect the concepts outlined in section 4.5, and to develop a systemic culture that acknowledges that this will take time.

In addition to self-directed and deep learning, the regulatory component of the model indicates that GenAI's future positive effects on students' learning will depend heavily on the point in the SRL cycle at which GenAI is employed. If GenAI provides the student with information before attempting the task - thereby eliminating forethought - GenAI will likely increase the learner's perception of their lack of agency. Conversely, if GenAI provides information after the attempt, thus allowing for performance and reflection, then it is likely to reduce that perception. Therefore, this creates a significant impact on how educational tools are developed and chosen: institutions considering the adoption of GenAI-based tools should consider not only the accuracy of the tool and its level of safety, but at what point during the completion of the task GenAI will enter by default, since that directly affects whether the tool encourages or discourages students from taking control over their learning (Lee et al., 2025).

Lastly, the framework's requirement for transparency has equity implications: institutions will no longer be able to purchase GenAI solely for efficiency once they are required to provide information about the origin of the data used in GenAI training and the inherent biases within GenAI in ways that learners can understand. Procurement will now become an exercise of epistemological governance. This will be particularly difficult for low-income districts that already experience limited access to GenAI due to both a digital divide and uneven AI literacy levels. Thus, GenAI may widen gaps that it intends to close (Pragya, 2025).

As mentioned previously, the framework gives policymakers a more precise definition of "using AI responsibly" than commonly found in many current institutional guidelines. In contrast to previous guidelines that refer to general principles related to transparency, fairness, and oversight, the framework's three dimensions and four boundary conditions translate into specific questions: has the division of labor in an assignment been clearly defined (relational)? Does the location of GenAI in a task promote or hinder the learning cycle (regulatory)? Is the school district prepared to make the investments necessary to move teachers along the integration continuum (pedagogical)? Are learners permitted to challenge or critique GenAI outputs? Will the institution reveal how GenAI was constructed (boundary condition)? While these questions will not have universally applicable answers (e.g., a First-Year Writing course and a doctoral research methods course will define labor differently), institutions must ask themselves these questions rather than gesturing toward 'balance'.

## 6. Limitations and Future Research

The study is theoretical, and there is still no empirical evidence on how using the Ecological Co-Agency Framework improves student achievement, equity among students, or the time teachers spend on their jobs. Although a significant amount of literature is reviewed in this paper, we conducted a targeted search (as described in Section 2) rather than a systematic one; we may have missed articles written in other languages, articles from regional journals, or articles from related fields.

This is not just an afterthought; A recent Special Section on the Cognitive and Epistemological Effects of Gen AI has specifically stated that an Interdisciplinary Research Agenda will be needed to measure both Efficiency and Epistemic Agency (Yan et al., 2025), and thus, the framework presented above is intended as a contribution to such a research agenda. Three Directions Follow: Firstly, studies demonstrating a relationship between each of the three dimensions of the framework (i.e., how clearly defined tasks are divided, at what point in the SRL Cycle an intervention occurs, and the extent to which institutions control the process), with measurable student achievement and/or learners' perception of autonomy. Secondly, a comparative study across different socio-economic contexts to assess whether the potentially inequitable impacts discussed in Section 5 occur in environments where resources are less available. Third- Longitudinal studies examining teachers' transition along the Pedagogical Co-Agent Continuum (Section 4.3) to determine if institutional investments or individual dispositions drive progress.

## 7. Conclusion

Generative AI is much more than just another classroom tool. Rather, when integrated into classrooms, generative AI reassigns agency and responsibility between humans and machines. Most of the literature (both technology-focused and technology-adoption-focused) does not capture this shift effectively. As stated above, one of the primary concerns associated with this shift of agency is epistemological in nature and therefore not primarily technical; the concern here is not that GenAI may produce unreliable content but that its ease of use may obscure whose judgment - either human or machine - ultimately underlies a particular assertion, determination, or product.

As such, the paper provided the development of the Ecological Co-Agency Framework. The framework views agency as relational, regulatory, and pedagogical, while a condition of human epistemic accountability constrains it. Specifically, the three conditions of the framework are: (1) contestability (the ability to question), (2) provenance (where did the information come from), and (3) no delegation of moral and/or intellectual responsibility. While the framework does not resolve the tensions between the potential pedagogical utility of AI and the problems discussed throughout this paper, it offers a way to address them through deliberation rather than chance. Therefore, the framework provides educators, institutions, and policymakers with a more precise language than the vague "balance" between human and artificial contributions, which is currently the common term used in many available discussions regarding AI. This intentionality will demand more than theoretical exploration alone. In addition to developing theoretical models based on the framework, researchers also plan to test it empirically across multiple educational settings.

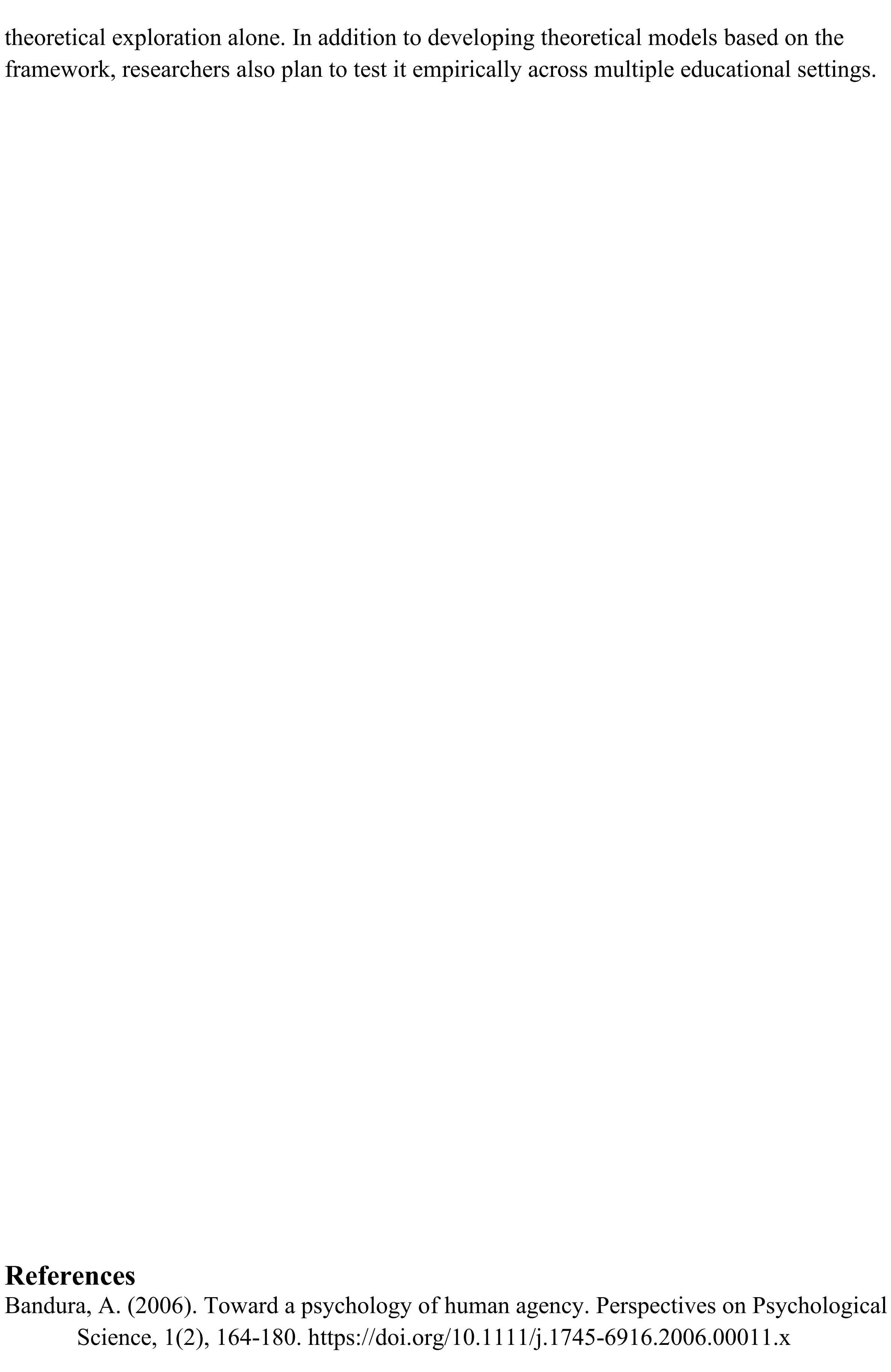